\documentstyle[preprint,aps]{revtex}

\begin{document}
\def\pt{\  .}
\def\com{\ ,}

\title{ THE INTERNAL GEOMETRY  OF AN  EVAPORATING BLACK  HOLE}

\author{Renaud Parentani and Tsvi Piran}

\address{Racah Institute of Physics
The hebrew University
Jerusalem, Israel 91904}
\maketitle

\begin{abstract}
We present a semi-classical model for the formation and evaporation of
a four dimensional black hole. We solve the equations numerically and
obtain solutions describing the entire the space-time geometry from
the collapse to the end of the evaporation. The solutions satisfy the
evaporation law: $\dot M \propto -M^{-2}$ which confirms dynamically
that black holes do evaporate thermally.  We find that the evaporation
process is in fact the shrinking of a throat that connects a
macroscopic interior ``universe" to the asymptotically flat
exterior. It ends either by pinching off the throat leaving a closed
universe and a Minkowskian exterior or by freezing up when the
throat's radius approaches a Planck size.  In either case the
macroscopic inner universe is the region where the information lost
during the evaporation process is hidden.
\end{abstract}


\centerline{\bf Introduction}

Hawking's discovery \cite{r2} of the thermal emission provoked by a
gravitational collapse has provided quantum mechanical grounds
\cite{r3} for the Bekenstein entropy \cite{r4}. Alas, it has also lead to an
apparent breakdown of predictability \cite{r1} precisely because of
the thermal character of the emitted flux.  In addition, if the black
hole completely evaporates, this breakdown of predictability results
in a loss of information since one is left with uncorrelated quanta
\cite{r5}. Confronted with this potential violation of the unitary evolution
of Quantum Mechanics, three scenarii have been proposed\cite{r21}: (i)
The loss of information, derived from the original semi-classical
treatment of Hawking is real \cite{r1,r6}.  (ii) The evaporation
process stops, leaving a Planckian remnant \cite{r7} which detains the
lost information.  (iii) A fully quantum description will reveal that
the information is properly recovered within the detailed structure of
the emitted quanta \cite{r8}.  In order to address these issues in a
simpler context, a two dimensional (2D) dilaton gravity model
\cite{r9} was introduced and analyzed intensively.
This model has indeed sharpened the problem \cite{r10,r11,r12} but has
not settled the question yet.

The resolution of this problem clearly requires a full dynamical
description of the evaporating black hole space-time. At present a
quantum mechanical description is beyond our reach. We present here a
four dimensional (4D) semi-classical model which describes the
formation and evaporation of a spherically symmetric black hole. We
solve the dynamical equations numerically and obtain the
space-time geometry till the Plancking regime where the approximation
breaks down and the final product depends on the regularization
scheme.  The solution confirms the generally accepted
features of black hole evaporation, which were anticipated assuming
the quasi stationarity character of the external geometry \cite{r18}.
But it also affords an explicit description of the interior geometry
and the emergence of the fluxes. In term of the external retarded
time, the interior region is static  because the outgoing
flux rises in the exterior part only.  Hence, at the end of the
evaporation one is left with a macroscopic interior region describing
a closed universe wherein the curvature is weak. Within the
semiclassical treatment, this closed region detains all the
correlations with the asymptotic Hawking quanta as well as the quantum
specification of the infalling matter.

\centerline{\bf The Model}

We use the spherical symmetric metric:
\begin{equation}
ds^2 =
e^{2f(U,V)} dU dV - r^2(U,V) d \Omega^2 \pt \label{e1}
\end{equation}
$U$ and $V$ are outgoing and ingoing null coordinates.  They are
specified
on the boundaries $U=U_-$ for $V>0$ and $V=0$ for $U>U_-$ by the
standard coordinate choice, $r=(V-U)/2$:
\begin{equation}
r(U,V=0)= -U/2 \ ; \ r(U_-,V)= (V-U_-)/2 \pt
\label{e5}
\end{equation}
$U_-$ is taken large and negative in order to approximate null
past infinity ($\cal I^-$).

We introduce two matter sources: a classical flux and a quantum field.
The first one is a radially infalling null dust ($T^{Cl}_{\mu\nu} =
\delta_\mu^V \delta_\nu^V F(V)/4\pi r^2$) carrying a mass $M_0$.
We choose $F(V)$ as a Gaussian centered around $V_i$ well inside our
$V$-domain.  Without a quantum backreaction, it would create an
eternal black hole of radius $2M_0$ if the width, $\sigma$ , is small
enough (i.e. $\sigma < M_0$) that the reflection at $r=0$ can be
neglected.  The second source is the expectation value of the energy
momentum tensor $T^{Q}_{\mu\nu}$.  In the absence of quantum on the
boundaries it is entirely determined by the geometry. This tensor
engenders, in turn, the Hawking flux and the black hole
evaporation. As a model, we take the expectation value of the tensor
\cite{r13} which arises from the quantization of a 2D massless field
in the 2D curved background described by $ ds^2 = e^{2f(U,V)} dU dV$
(see Eq. \ref{e1}) divided by $4 \pi r^2$ to mimic the four
dimensional radial dependence. It reads simply:
\begin{equation}
T^{Q}_{UV} = {-\alpha f_{,UV}  \over 4 \pi r^2}
\com
\label{e2}
\end{equation}
\begin{equation}
T^{Q}_{UU} = {\alpha [f_{,UU}-(f_{,U})^2]
\over 4 \pi r^2}
\ ;  \
T^{Q}_{VV} = {\alpha [f_{,VV}-(f_{,V})^2]  \over 4 \pi r^2}
\pt
\end{equation}
The normalization constant $\alpha$ depends linearly on the number of
massless fields and will control the rate of evaporation (see
Eq.~\ref{e7} below).  In 2D, this expectation value (times $4 \pi
r^2$) gives an asymptotic flux which coincides with the flux obtained
from the Bogoliubov transformation \cite{r14} relating the scattered
modes to the asymptotic ones.  In 4D, it offers a good approximation
since most of the energy is carried away by s-waves quanta \cite{r15}.

In the metric Eq. \ref{e1}, the Einstein equations split into two
dynamical equations and two constraints.  The classical source
$T_{\mu\nu}^{Cl}$ appears in one constraint only while $T_{UV}^{Q}$
modifies the dynamical Einstein equations.  On the left of the
infalling matter, for $V\ll V_i$, one has Minkowski space-time.
Indeed, on $V=0$, the first constraint gives $\partial f(U,0)/\partial
U =0$. Then by choosing $f(U,0)=0$, one equates $(U+V)/2$ to the
proper time at rest with respect to the spherical infalling shell.  On
$U=U_-$, the second constraint gives
\begin{equation}
f_{,V}(U_- ,V) = 8 \pi r (T_{VV}^{Cl})
\pt  \label{e6}
\end{equation}
A simple integration of Eq.~\ref{e6} gives the Cauchy date for the
dynamical integration, $f(U_- ,V)$.

The modified dynamical equations read:
\begin{equation}
f_{,UV} (1-\alpha e^{2d}) =
d_{,U} d_{,V} +  e^{2(d+f)} /4
\label{e3}
\end{equation}
\begin{equation}
d_{,UV} (1-e^{2d}) = d_{,U} d_{,V}
(2-\alpha e^{2d}) + e^{2(d+f)} /4 \com
\label{e4}
\end{equation}
where we define $d \equiv -\ln (r)$.  The inclusion of the quantum
matter term (Eq. \ref{e2}) leads, as in dilatonic gravity \cite{r9},
to singular equations at the critical radius $r_\alpha =
\sqrt{1/\alpha}$. In 4D this singularity is an artifact of our model
which assumes the validity of expectation values (Eq. \ref{e2}) in the
Plankian domain. Still this method offers \cite{r16} a reliable
approximation during most of the evaporation process when the
prefactor $P=1-\alpha e^{2d} $ is still close to $1$. If one wishes to
carry out the integration into the Planckian domain one can eliminate
this singularity by regularizing the prefactor $P$ at small radii in
the following way:
\begin{equation}
P_n = 1-{\alpha  e^{2d} \over 1+ (\alpha  e^{2d})^n }
\pt \label{e10xx}
\end{equation}
The final stage of the evaporation, where the mass approaches the
Plank mass, will depend then on $n$.

\centerline{\bf  Geometrodynamics }

We integrate numerically the dynamical equations using the Cauchy data
and obtain the geometry of the collapse and the subsequent
evaporation.  The evaporation stage is contained in a tiny $U$ lapse
(Recall that the ($U,V$) coordinates constitute the inertial frame of
an observer at rest at the origin).  We make, therefore, a coordinate
transformation to ($u,v$), the inertial asymptotic system
at rest with respect to the black hole. $(u,v)$ are defined by $dr
/du_{|v} \rightarrow -1/2$, $dr/dv_{|u} \rightarrow 1/2$ for $r
\rightarrow \infty$.   With this transformation the   tiny $U$ lapse
is spread out since any inertial collapse results in a Doppler shift
given by:
\begin{equation}
{dU \over du} = e^{-u/4M_0}
\label{ev}
\end{equation}
wherein the imaginary frequency determines the initial Hawking
temperature $1 /8 \pi M_0$.

Fig. 1 depicts the geometry.  On the left on finds the inner boundary
of the trapped region (defined by $\partial r/\partial v=0$). It is
space-like because of the positivity of the classical infalling energy
\cite{r17}.  This boundary starts at $r=0$ (outside of the figure) and
ends when the negative energy flux, that causes the evaporation,
equals the tail of the classical flux (at $ u=208$).  The outer part
of the trapped region is the apparent horizon, $r_{ah}$.  Since the
infalling quantum flux is negative, $r_{ah}(v)$ is time-like
\cite{r18}.  It ends upon reaching (at $ v=345$) the critical radius
$r_\alpha$ (not drawn) at which our dynamical equations break down.

We note that the apparent horizon is almost a ``static'' line: $v = u
+ const$. This proves that the inwards flux ($T_{vv}$) on this horizon
is equal and opposite to the outgoing flux ($T_{uu}$) at infinity.
The mass loss can be calculated from either flux.  In practice we
measure it from $dr_{ah} /dv$ using $ r_{ah}=2M_{ah}$.  When the mass
is still large (compared to the critical mass $M_\alpha=2 r_\alpha$)
our dynamical model yields:
\begin{equation}
{d M_{ah} \over dv} = {- \alpha \over 32} { 1 \over M_{ah}^2 (v) }
\pt \label{e7}
\end{equation}
This demonstrates that ``black holes do evaporate thermally'' in
perfect agreement with Bardeen \cite {r18} whose original
derivation  assumes a quasi stationary character of the external
geometry.  Naturally, when the mass approaches $M_\alpha$ the mass
loss depends on the regularization scheme (see Eq.~\ref{e10xx} and
Fig. 2).  For $n=1$ we find that $\dot M$ increases sharply while for
$n=2$, a rapid drop in $\dot M$ and in the temperature appears
upon approaching the critical mass.

In order to illustrate the internal geometry we have depicted, in
Fig. 3, the radii of the spheres encountered by successive outgoing
light rays (i.e. along $u=const$) in the Eddington-Finkelstein
coordinates $(r,v)$. When the matter is far away from its
Schwarzschild radius, the evolution begins with $ r_u(v) = (v-u)/2$.
In the absence of backreaction it would have ended with the asymptotic
line which  starts at $r=0$ and  approaches asymptotically
$r=2M_0$, forming the event horizon.  When back reaction is included
the null rays still spread out from $r=0$ to $r=2M_0$ but then
contract within the trapped region reaching a minimal radius at $
r_{ah}$ before spreading out again.  As the black hole losses it mass
$r_{ah}$ diminishes accordingly.  This geometry describes a shrinking
``throat" that separates the inner ``universe" around the infalling
mass from the exterior space-time.  Since the inner universe is static
(see Fig 1b) during the whole evaporation, it remains macroscopic.
Thus, what seems to be, for an external observer, the evaporation of
the black hole, is in fact the shrinking of the throat that connects
the internal region to the rest of the world \cite{r19}.

\centerline{\bf  Implications for the Loss of Information}

Before analyzing what implications to the loss of information paradox
this geometry might give, we recall that the geometry was obtained
using a semi-classical approximation.  Thus, the forthcoming analysis
is meaningful only if the {\it mean} geometry is correctly
approximated by the semi-classical method (beware that, for instance,
the ``trans-Planckian'' fluctuations \cite{r16} might completely ruin
it).  Then, in order to address the problem of information loss, which
deals with quantum correlations, one has to analyze the quantum
fluctuations on the resulting background.

Before the pinching off the Schrodingerian evolution of the quantum
matter state on all our slices \cite{r20} can be used to evaluate the
correlations between the field configurations inside and outside the
throat.  As the evaporation approaches the Planckian regime, a region
of high curvature appears {\it only} at the throat, separating
``practically" the left unaffected interior region from the exterior
space-time. Hence whether or not the evaporation stops, does not
change significantly the situation.  If the black hole completely
disappears, one finds two disconnected macroscopic regions: a quite
big ``baby universe" \cite{r19} and a Minkowskian exterior.  On the
other hand, if a remnant characterized by a Planckian size throat is
left, there is still a tiny connection with the macroscopic internal
region.  In either case, the quantum matter field configurations in
the internal region are still correlated to external configurations as
they were just before the pinching off. The loss of information in the
exterior part of the space-time is analogous \cite{r6} to the loss of
the quantum correlations which occurs in any subsystem upon tracing
over the quantum states belonging to its complementary subsystem.  The
``new'' behavior that appears to an outside observer occurs because
The dynamics of the evaporation force quantum mechanics to operate in
the realm of space-time with a varying topology.

This proposal requires, nevertheless, a drastic modification of the
interpretation of the Bekenstein entropy. The area of the black hole
does not count the (log of the) number a microscopic states contained
in the internal region but only the number of them which are still
coupled to the external world.  (For instance, this entropy reflects
the potential increase in the external entropy due to a complete
evaporation of the black hole and has to be used upon enclosing the
black hole in a cavity and searching for thermodynamical equilibrium
since only the states coupled to the external world states could
participate to the thermalisation.)  The universality of this
area-entropy comes now from the universality of the characteristics of
the throat (the throat has no hair) that connects the interior region
to the rest of the world.  Within the interior region the number of
available micro-states depends explicitly on the history of the
collapse since the characteristic size of this region has nothing to
do with the actual throat radius.

We would like to thank J. Bekenstein for many illuminating
conversations.  The research was funded by a grant from the US-Israel
Binational Science Foundation to the Hebrew University.


\figure{ Fig. 1a:}
{\it Contour lines of $r=const$ from $r=3$ (on the upper left corner)
to $r=80$ (in the lower right corner).  The heavy line depicts the
apparent horizon $r_{ah} $.  The infalling matter is centered
initially at $v_i=14$ and it has a width of $\sigma=4.5$. For $v \ll
v_i$ the space-time is Minkowski.  For $v\gg v_i$ it has a
Schwarzchild geometry characterized by the time dependent residual
black hole mass.

\figure{Fig 1b:}
{\it Contour lines of $f=const$ of the same evaporating geometry.  The
heavy dashed line designates the locus where $T^{Q}_{uu}$ reaches half of its
asymptotic value.  It coincides  with the places where $
f_{,VU}$ is maximum and it is always outside the trapped region. The
fact that $ f_{,VU}$ vanishes inside the trapped region indicates that
$T^{Q}_{vv}$ is constant and $T^{Q}_{uu}$ is zero. This implies that
the infalling matter keeps collapsing under its own gravitational
field ignoring completely the Hawking flux which develops outside its
past light-cone.  }

\figure{Fig. 2.} The logarithm of the rate of evaporation
vs. the logarithm of the residual mass for different regularization
schemes (see Eq. \ref{e10xx}): $n=1$ (solid line) with an exploding
solution and $n=2$ (dashed line) with a decreasing evaporation rate.
The non regularized solution ($n=0$) follows the $n=2$ curve and stops
at the critical mass around the intersection of the curves. Note that
the slope, being -2 gives Eq. \ref{e7}.

\figure{Fig. 3:}
{\it   The radii of spheres ($r_u (v)$) on successive
$u=const$ slices in advanced Eddington-Finkelstein coordinates $(r,v)$.
The bold line denotes the center of the infalling matter $v=v_i$.

\begin{thebibliography}{99}

\bibitem{r2}S.W. Hawking, Comm. Math. Phys. {\bf 43} (1975)
199.

\bibitem {r3}  S.W. Hawking, Phys. Rev.  {\bf D13},  (1976) 191.

\bibitem{r4}J.D. Bekenstein, Phys. Rev. {\bf D9} (1974)
3292; Phys. Rev. {\bf D7} (1973) 2333.

\bibitem{r1}S.W. Hawking, Phys. Rev. {\bf D14} (1976) 2460.

\bibitem{r5}R.M. Wald, Comm. Math. Phys. {\bf 45} (1975) 9.

\bibitem{r21}  For a review see: J. Preskill  CALT-68-1819, hep-th/9209058
Presented at International Symposium on Black holes, Membranes, Wormholes and
Superstrings, Woodlands, TX, 16-18 Jan 1992.


\bibitem{r6}R.M. Wald, in: Black Hole Physics. Erice
Lectures 1991. NATO ASI Series, Kluwer Academic
Publishers, 1992.


\bibitem{r7} c.f. Ref.1 and
Y. Aharonov, A. Casher and S. Nussinov, Phys. Lett {\bf 191B} (1987)
51.

\bibitem{r8}D.N. Page, Phys. Rev. Lett. {\bf 44} (1980) 301.

\bibitem{r9}C.G. Callan, S.B. Giddings, J.A. Harvey and A.
Strominger, Phys. Rev. {\bf D45} (1992) 1005;
for a review: J.A. Harvey and A. Strominger, hep-th/9209055

\bibitem{r10}   J.G. Russo, L. Susskind and L. Thorlacius, Phys.
Rev. {\bf D47} (1993) 533.

\bibitem{r11}  T.Piran and A.Strominger, Phys. Rev. {\bf D48}  (1993) 4729.
\bibitem{r12}  K. Schoutens, E. Verlinde and H. Verlinde, Phys. Rev. {\bf D48}
(1993) 2670

\bibitem{r18}  J.  Bardeen, Phys. Rev. Lett.{\bf 46 }, (1981) 382 .

\bibitem{r13}P.C.W. Davies, S.A. Fulling and W.G. Unruh,
Phys. Rev. {\bf D13} (1976) 2720; See
also S.M. Christensen and S.A. Fulling, Phys. Rev. {\bf D15} (1977) 2088.

\bibitem{r14}  N.D. Birrel and P.C.W. Davies, Quantum Field in Curved Space,
Cambridge University Press (1982); S. Massar, R. Parentani and
R. Brout, Class. Quant. Grav. {\bf 10} (1993) 2431.

\bibitem{r15}  P. Candelas, Phys. Rev. {\bf D21}, (1980) 2185;
D. Page, Phys. Rev. {\bf D13}, (1976) 198.

\bibitem{r16} T. Jacobson, Phys. Rev. {\bf D48}  (1993) 728;
R. Stephens, G. 't Hooft and B.F. Whiting, THU-93/20; UF-RAP-93-11,
gr-qc/9310006 and F. Englert, S. Massar and R. Parentani,
gr-qc/9404026 have questioned recently this approximation even when
the curvature and the flux are very far from the Planckian regime.
The basic reason being that the frequencies involved in the
fluctuations around the mean fluxes are trans-Planckian.

\bibitem{r17}  S. W. Hawking, and G. F. R. Ellis {\it The Large Scale
Structure of Space-Time}, Cambridge University Press, Cambridge, England,
1973.

\bibitem{r19}  S. W. Hawking, R. Laflamme, Phys, Lett {\bf B209},  (1988)  39.

\bibitem{r20}  There is no need for a
``super-observer" discussed by J.G. Russo, L. Susskind and
L. Thorlacius, Phys. Rev. {\bf D49} (1994) 966 on our slices for
revealing the detailed correlations. All the information stored in
them can, in principle, be communicated to an asymptotic observer.






\end{thebibliography}
\end{document}